\documentclass[preprint,aps,prd]{revtex4}
\pdfoutput=1

\usepackage{braket}             % Allows Dirac notation
\usepackage{graphicx}           % Include figure files
\usepackage{bm}             % bold math
\usepackage[usenames, dvipsnames]{color}
\usepackage{hyperref}           % add hypertext capabilities
\usepackage{mathtools}
\usepackage[normalem]{ulem} %strikethrough text
\usepackage[caption=false]{subfig}
\usepackage{float}
\usepackage{graphicx}
\usepackage{epstopdf}
\DeclareMathOperator{\tr}{tr}           % for trace
\usepackage{colonequals} %Get a nice looking :=

\usepackage[usenames, dvipsnames]{xcolor}

\definecolor{capri}{rgb}{0.0, 0.75, 1.0}

\begin{document}

\title{Harvesting Entanglement by non-identical detectors with different energy gaps}

\author{Hui Hu, Jialin Zhang~\footnote{Corresponding author at jialinzhang@hunnu.edu.cn} and Hongwei Yu~\footnote{Corresponding author at hwyu@hunnu.edu.cn}}
\affiliation{Department of Physics and Synergetic Innovation Center for Quantum Effects and Applications, Hunan Normal University, Changsha, Hunan 410081, China}
\date{\today}

\begin{abstract}
It has been shown that the vacuum state of a free quantum field is entangled and such vacuum entanglement  can be harvested by a pair of initially uncorrelated  detectors interacting locally with the vacuum field for a finite time. In this paper, we examine the entanglement harvesting phenomenon of two non-identical inertial detectors with different energy gaps locally interacting with massless scalar fields via a Gaussian switching function.  We focus on how entanglement harvesting depends on the energy gap difference  from two perspectives: the amount of entanglement harvested  and the harvesting-achievable separation between the two detectors. In the sense of the amount of entanglement, we find that as long as the inter-detector separation is  not too small with respect to the interaction duration parameter, two non-identical detectors could extract more entanglement from the vacuum state than the identical detectors. There exists an optimal value of the energy gap difference when the inter-detector separation is sufficiently large  that  renders the harvested entanglement to peak. Regarding the harvesting-achievable separation,  we further find that the presence of an energy gap difference  generally enlarges the harvesting-achievable separation range. Our results  suggest that  the non-identical detectors may be advantageous
to extracting entanglement from vacuum in certain circumstances as compared to identical detectors.
\end{abstract}

\maketitle

%========================================
%========================================

\section{Introduction}
 Quantum entanglement  is an intriguing phenomenon  in  quantum physics and  a crucial resource in  quantum information technologies, e.g., quantum communication~\cite{RC,Bh,HorodeckiHHH:2009}, quantum teleportation~\cite{Bg}, dense coding~\cite{Bh} and quantum key distribution~\cite{NSSPLBU:2021}. In recent years, a lot of progress has been made in understanding entanglement in  quantum field  theory. It has been demonstrated that the vacuum state of a free quantum field is entangled in the framework of the formal
algebraic quantum field theory~\cite{SummersW:1985,Summers:1987}, in the sense that the vacuum state can maximally violate  Bell's inequalities. Interestingly, such vacuum entanglement  was found to be able to even be extracted by a pair of initially uncorrelated particle detectors/atoms interacting locally with vacuum fields for a finite time~\cite{Valentini:1991,Reznik:2003,Reznik:2005}. More recently, this extraction process has been further operationalized in a protocol with the Unruh-DeWitt (UDW)  detector model of two levels~\cite{Martin-Martinez:2013eda,Salton-Man:2015}, and the phenomenon has  been known as entanglement harvesting~\cite{Salton-Man:2015,Pozas-Kerstjens:2015}.

The entanglement harvesting phenomenon has attracted a lot
of attention and  been examined in a wide variety of circumstances, e.g., with detectors in accelerated motion, in the presence of completely reflecting boundaries  and in spacetimes with nontrivial topology. It was demonstrated that the acceleration of detectors, which leads to  the well-known Unruh effect,
can surprisingly assist entanglement harvesting in certain cases\cite{Salton-Man:2015,Zhjl:2021-2} and the reflecting boundary plays a double edged
role in entanglement harvesting, i.e., degrading in general the harvested entanglement
while enlarging the entanglement harvesting-achievable parameter space~\cite{Zhjl:2021}.  More intriguingly, one also  finds that entanglement harvesting is quite sensitive to
the intrinsic  properties of spacetime including its dimension~\cite{Pozas-Kerstjens:2015}, topology~\cite{MST:2016}, curvature~\cite{Steeg:2009,Kukita:2017,Zhjl:2019,Robbins:2020jca,Ng:2018-2,Zhjl:2018},
and causal structure~\cite{Henderson:2020zax,Tjoa:2020,Gallock-Yoshimura:2021}, and thus the phenomenon of entanglement harvesting  may provide a tangible  way to probe the properties of spacetime ~\cite{MST:2016,Kukita:2017,Ng:2018-2,Zhjl:2018} and even distinguish a thermal bath from an expanding universe at the same temperature~\cite{Steeg:2009}. Experimentally, some feasible setups for entanglement harvesting,  such as atomic  systems, superconducting circuits and a microcavity,
 have been proposed in several exploratory studies~\cite{Olson:2011,SDRGL:2011,SPDRM:2012,Ardenghi:2018,Beny:2018}.

So far, the study of entanglement harvesting  is  focused on the situation of two detectors with an equal energy gap, i.e., two identical detectors,  whereas  little is known for the case of non-identical detectors.
Actually, according to  the entanglement harvesting protocol, an energy gap difference  between two  detectors manifests in both  detectors' transition probabilities and the nonlocal correlations of the quantum field in vacuum  which in turn jointly determine the  entanglement harvesting. Hence, it is quite interesting and sensible to explore entanglement harvesting for  non-identical detectors and see whether  novel properties emerge.  %Motivated by these above,
In this paper,  we  present such a study on entanglement
harvesting for two non-identical detectors and address the question whether  non-identical detectors could harvest more entanglement than identical ones with the entanglement harvesting protocol and how an  energy gap difference between the two detectors  affects the harvesting phenomenon.

The rest of this paper is organized as follows. We begin in Section II by reviewing the basic formalism for the UDW detectors locally interacting with vacuum scalar fields in the framework of the entanglement harvesting protocol. In Section III, we analyze the properties of entanglement harvesting for two inertial detectors with different energy gaps, and make a cross-comparison with that for two identical ones with an equal energy gap. %We also  focus on which configuration of non-identical detectors is the optimal setting for  entanglement harvesting.
Finally, we finish with conclusions in Section IV.  Throughout the paper the natural units $\hbar=c=1$ are  adopted.

\section{The basic formulas}
 We consider a pair of point-like two-level static detectors  (labeled by $A$ and $B$ )   with $|g\rangle$ and $|e\rangle$  denoting  the ground  and excited states respectively, which locally  interact with  a massless scalar field $\phi$  in vacuum, and assume that the  interaction Hamiltonian takes the following  form in the interaction picture
 \begin{equation}\label{eqt1}
   H_{D}(\tau)=\lambda \chi(\tau)[e^{i\Omega_{D}\tau}\sigma^{+}+e^{-i\Omega_{D}\tau}\sigma^{-}]\phi[x_{D}(\tau)],
   \;\;D\in \{A,B\}
   \end{equation}
where $\lambda$ is the coupling strength and  subscript $D$ specifies  which detector we are considering, $\chi(\tau):=\exp[-\tau^{2}/(2\sigma^{2})]$ is a switching function with  parameter $\sigma$ controlling the duration of interaction,  $\Omega_D$ is the energy gap of the detector,   $\sigma^{+}=|e\rangle_{D} \langle g|_{D}$, $\sigma^{-}=|g\rangle_{D} \langle e|_{D}$ are the $SU(2)$ ladder operators acting on the Hilbert space of  the detector, and $x_{D}(\tau)$ is the spacetime  trajectory of the detector parameterized by its proper time $\tau$.

Initially, we suppose that the two UDW detectors are prepared in their ground state and the field is in the vacuum
state $|0\rangle_{M}$. With the interaction Hamiltonian Eq.~(\ref{eqt1}), the final reduced density matrix of the
system (two detectors) can be obtained in the basis  \{$|g\rangle_{A}|g\rangle_{B},|g\rangle_{A}|e\rangle_{B},
|e\rangle_{A}|g\rangle_{B},|e\rangle_{A}|e\rangle_{B}$\},  after  some algebraic manipulations based on the perturbation theory~\cite{Pozas-Kerstjens:2015,MST:2016,Zhjl:2018}
 \begin{align}\label{rhoAB}
 \rho_{AB}:&=\tr_{\phi}\big(U \ket{\Psi}_{i}\bra{\Psi}_{i} U^{\dag}\big)\nonumber\\
 &=\begin{pmatrix}
 1-P_A-P_B & 0 & 0 & X \\
 0 & P_B & C & 0 \\
 0 & C^* & P_A & 0 \\
 X^* & 0 & 0 & 0 \\
 \end{pmatrix}+{\mathcal{O}}(\lambda^4)\;,
 \end{align}
 where
     \begin{equation}\label{PAPB}
 P_D:=\lambda^{2}\iint d\tau d\tau' \chi(\tau) \chi(\tau') e^{-i \Omega_{D}(\tau-\tau')}
 W\left(x_D(\tau), x_D(\tau')\right)\quad\quad D\in\{A, B\}\;,
\end{equation}
     \begin{align}
 C &:=\lambda^{2} \iint d \tau d \tau^{\prime} \chi(\tau) \chi(\tau')
 e^{-i (\Omega_{A}\tau-\Omega_{B}\tau')} W\left(x_{A}(\tau), x_{B}(\tau')\right)\;,
 \end{align}
 \begin{align}\label{xxdef}
 X:=&-\lambda^{2} \iint d\tau d \tau' \chi(\tau)\chi(\tau') e^{-i(\Omega_{A}\tau+\Omega_{B}\tau')}
 \Big[\theta(\tau'-\tau)W\left(x_A(\tau),x_B(\tau')\right)\nonumber\\&+\theta(\tau-\tau')W\left(x_B(\tau'),x_A(\tau)\right)\Big]\;
 \end{align}
with $W(x,x'):=\bra{0}_M\phi(x)\phi(x')\ket{0}_M$ denoting  the
Wightman function of the field   and  $\theta(\tau)$ the Heaviside step function. Here, we have used the fact that  $t=\tau$ for detectors at rest.
Let us note that $P_{D}$ is the transition probability of detector $D$, and the quantities $C$ and $X$  characterize the field correlations~\cite{Zhjl:2018}.  We  utilize concurrence  as a measure of  entanglement harvested by the UDW detectors, which can be evaluated straightforwardly from  the $X$-type density matrix  Eq.~(\ref{rhoAB}) to give~\cite{MST:2016,Zhjl:2018,Zhjl:2019}
\begin{equation}\label{condef}
\mathcal{C}(\rho_{A B})=2 \max \Big[0,|X|-\sqrt{P_{A}
P_{B}}\Big]+\mathcal{O}(\lambda^{4})\;.
\end{equation}
So, the concurrence quantifying the harvested entanglement is  a  result of the competition between off-diagonal matrix element $X$ and the geometric mean of transition probabilities $P_A$ and $P_B$. For two identical detectors at rest,  $P_A=P_B$, and the concurrence  is simply determined by  $|X|-P_{D}$. %Therefore, obtaining the exact functional expression of $X$ and $P_D$  in a variety of scenarios  is a crucial step to gain a deep insight as to  entanglement harvesting.

 %%%%%%%%%%%%%%this relevant quantity signifying the concurrence is given by

\section{  Entanglement harvesting for non-identical detectors}
We now examine entanglement harvesting for two  static non-identical detectors with different energy gaps.  Without loss of generality, we assume the detector labelled by $A$  has a comparatively smaller energy gap, i.e., $\Delta\Omega:=\Omega_B-\Omega_A\geq0$ throughout the paper.
In  the four dimensional Minkowski
spacetime, the Wightman function for the massless scalar field is given by~\cite{Birrell:1984}
\begin{equation}\label{wightdef}
W(x,x')=-\frac{1}{4\pi^2}\frac{1}{(\tau-\tau'-i\epsilon)^2-|\mathbf{x}-\mathbf{x'}|^{2}}\;.
\end{equation}
Substituting  Eq.~(\ref{wightdef}) into Eq.~(\ref{PAPB}), one can find the transition probability~\cite{Zhjl:2021,MST:2016}
 \begin{equation}\label{PDExpression}
 P_{D}=\frac{\lambda^{2}}{4\pi}\Big[e^{-\sigma^{2}\Omega_D^{2}}-\sqrt{\pi}\Omega_D\sigma
{\rm{Erfc}}(\sigma\Omega_D)\Big],
~~ D\in\{A,B\}\;,
\end{equation}
where ${\rm{Erfc}}(x):=1-{\rm{Erf}}(x)$ with the error function ${\rm{Erf}}(x):=\int_0^x2e^{-t^2}dt/\sqrt{\pi}$.

We suppose that two detectors are separated by a distance $L$,  then the  correlation term  $X$  can be straightforwardly worked out from Eq.~(\ref{xxdef})(see Appendix.\ref{appd1})
\begin{align}\label{Expression-X}
    X&=-\frac{\lambda^{2}\sigma}{8\sqrt{\pi}L}
    e^{-\frac{\sigma^{4}(2\Omega_A+\Delta\Omega)^{2}+L^{2}}{4\sigma^{2}}}
    \;\Big[e^{i\Delta\Omega{L}/2}
    {\rm{Erfi}}\Big(\frac{L-i\sigma^2\Delta\Omega}{2\sigma}\Big)\nonumber\\
    &+e^{-i\Delta\Omega{L}/2}{\rm{Erfi}}\Big(\frac{L+i\sigma^2\Delta\Omega}{2\sigma}\Big)
    +2i\cos\Big(\frac{\Delta\Omega{L}}{2}\Big)\Big]\;,
\end{align}
where the imaginary error function is defined as ${\rm{Erfi}}(x):=-i{\rm{Erf}}(ix)$. Let us note here that the correlation term $X$  diverges in the limit of $L\rightarrow0$.  This divergence actually arises from the ill-defined  point-like  approximation of the UDW detector model in the entanglement harvesting protocol  when $L/\sigma\ll\lambda$, and a finite-size detector model with a spatial smearing function should be called for to resolve the issue~\cite{Pozas-Kerstjens:2015}.

Although further analytical behaviors of the  transition probabilities $P_D$ and the  correlation term $X$  are not obtainable, approximate expressions can however be found in some special cases.
%here can be written in analytical forms, the detailed behavior of  harvested entanglement (concurrence) is still not easy to capture due to its complicated functional expression.  However,
For energy gaps extremely small as compared to the duration of interaction parameter $\sigma$ ($\Omega_A\sigma\leq\Omega_B\sigma\ll1$ ), we have
\begin{equation}\label{app-papb1}
\sqrt{P_AP_B}\approx\frac{\lambda^{2}}{4\pi}e^{-\sigma^{2}[\Omega_A^{2}+(\Omega_A+\Delta\Omega)^2]/2}\;,
\end{equation}
whereas for large energy gaps ($1\ll\Omega_A\sigma\leq\Omega_B\sigma$),
\begin{equation}\label{app-papb2}
\sqrt{P_AP_B}\approx\frac{\lambda^{2}}{8\pi}\frac{e^{-\sigma^{2}[\Omega_A^{2}+(\Omega_A+\Delta\Omega)^2]/2}}{\Omega_A(\Omega_A+\Delta\Omega)\sigma^2}\Big[1-\frac{3}{4\sigma^2\Omega_A^2}
-\frac{3}{4\sigma^2(\Omega_A+\Delta\Omega)^2}\Big]\;.
\end{equation}
Regarding the correlation term~(\ref{Expression-X}), we find that  it can be approximated,  when  $L/\sigma\ll 1$,  as
\begin{equation}\label{app-X}
X\approx-\frac{\lambda^{2}{e}^{-{\sigma^{2}(2\Omega_A+\Delta\Omega)^{2}}/4}}{4\sqrt{\pi}}
    \Big[\frac{i\sigma}{L}+\frac{e^{-\Delta\Omega^2\sigma^2/4}}{\sqrt{\pi}}+\frac{\Delta\Omega\sigma}{2}\rm{Erf}\Big(\frac{\Delta\Omega\sigma}{2}\Big)\Big]\;,
\end{equation}
while  $L/\sigma\gg 1$, the approximation is
\begin{equation}\label{app-X2}
X\approx-\frac{\lambda^{2}\sigma}{4\sqrt{\pi}L}\Big[
\frac{2\sigma}{L\sqrt{\pi}}{e}^{-\sigma^{2}[\Omega_A^{2}+(\Omega_A+\Delta\Omega)^2]/2}
+ie^{-\frac{\sigma^{4}(2\Omega_A+\Delta\Omega)^{2}+L^{2}}{4\sigma^{2}}}\cos\Big(\frac{L\Delta\Omega}{2}\Big)\Big]\;.
\end{equation}
Therefore, one can estimate the concurrence for small  inter-detector separations ($L/\sigma\ll 1$) to be
\begin{equation}\label{app-c1}
\mathcal{C}(\rho_{A B})\approx\frac{\lambda^{2}\sigma}{2L\sqrt{\pi}}{e}^{-{\sigma^{2}(2\Omega_A+\Delta\Omega)^{2}}/4}\;,
\end{equation}
while that for large inter-detector separations (${L/\sigma}\gg1$)
\begin{equation}\label{app-c2}
\mathcal{C}(\rho_{A B})\approx\left\{\begin{aligned}
&\max \Big\{0, \frac{\lambda^{2}}{2\pi}e^{-\sigma^{2}[\Omega_A^{2}+(\Omega_A+\Delta\Omega)^2]/2}
\Big(\frac{2\sigma^2}{L^2}-1\Big)\Big\}=0,&\Omega_A\sigma\leq\Omega_B\sigma\ll1;\\
      & \max \Big\{0, \frac{\lambda^{2}}{2\pi}e^{-\sigma^{2}[\Omega_A^{2}+(\Omega_A+\Delta\Omega)^2]/2}
\Big[\frac{2\sigma^2}{L^2}-\frac{1}{2\Omega_A(\Omega_A+\Delta\Omega)\sigma^2}\Big]\Big\},&\Omega_B\sigma\geq\Omega_A\sigma\gg1\;.
 \end{aligned} \right.
\end{equation}
As  can be seen from Eq.~(\ref{app-c1}), the entanglement harvested by time-like separated detectors ($L/\sigma\ll1$) is an  obviously decreasing function of the energy gap difference, while it follows from Eq.~(\ref{app-c2}) that for  space-like separated detectors ($L/\sigma\gg1$)   entanglement harvesting is only possible for detectors with an energy gap larger than  the
 Heisenberg energy (i.e, $\Omega_B\geq\Omega_A\gg1/\sigma$ ). This can be understood as follows.

  For a very small inter-detector separation($L/\sigma\ll1$),  one can see from Eq.~(\ref{condef}) and Eq.~(\ref{app-X}) that  the correlation term $X$ will
be a decisive player in the concurrence and the term related to the transition probabilities can be neglected, leading to  %and the leading term of $X$  renders
that the concurrence behaves as a decreasing function of the energy gap difference.
However, for a large inter-detector separation ($L/\sigma\gg1$), the term related to the transition  probabilities  can no longer be  neglected because now the correlation term $X$ for two space-like  separated detectors also becomes small. For detectors with an energy gap much smaller than the  Heisenberg energy (i.e, $\Omega_A\ll1/\sigma$ ), the probabilities term exceeds the correlation term, thus making the concurrence vanish, whereas for detectors with  an energy gap much larger than the  Heisenberg energy ($\Omega_A\gg1/\sigma$),  the probabilities term can be smaller than the correlation term $X$  as long as the inter-detector separation is not too large, since  the detector's transition probability is now exponentially suppressed, yielding a non-zero concurrence.

  Since the concurrence is determined by the competition between the correlation $X$ and the transition probabilities, one may infer that  for  any detectors with  given energy gaps  the  entanglement will not be able to be harvested for a very  large but finite inter-detector separation. In other words, there exists a   separation range for  entanglement harvesting to be possible. For convenience, we  introduce  parameter $L_{\rm{max}}$  to characterize
the maximum harvesting-achievable separation beyond which entanglement harvesting ceases to occur.  It easy to deduce from Eq.~(\ref{app-c2}) that $L_{\rm{max}}$, for very large energy gaps ($\Omega_A\sigma\gg1$), takes the form
\begin{equation}
L_{\rm{max}}\approx2\sigma^2\sqrt{\Omega_A(\Omega_A+\Delta\Omega)}\;,
\end{equation}
and this reveals that  an energy gap difference can  enlarge the harvesting-achievable range.

In order to show the  behavior of the entanglement harvested by non-identical detectors in more general cases,  we will now perform numerical evaluations on the concurrence.
\begin{figure}[!ht]
\centering
\subfloat[$\Omega_A\sigma=0.50$]{\label{con-L11}\includegraphics[width=0.47\linewidth]{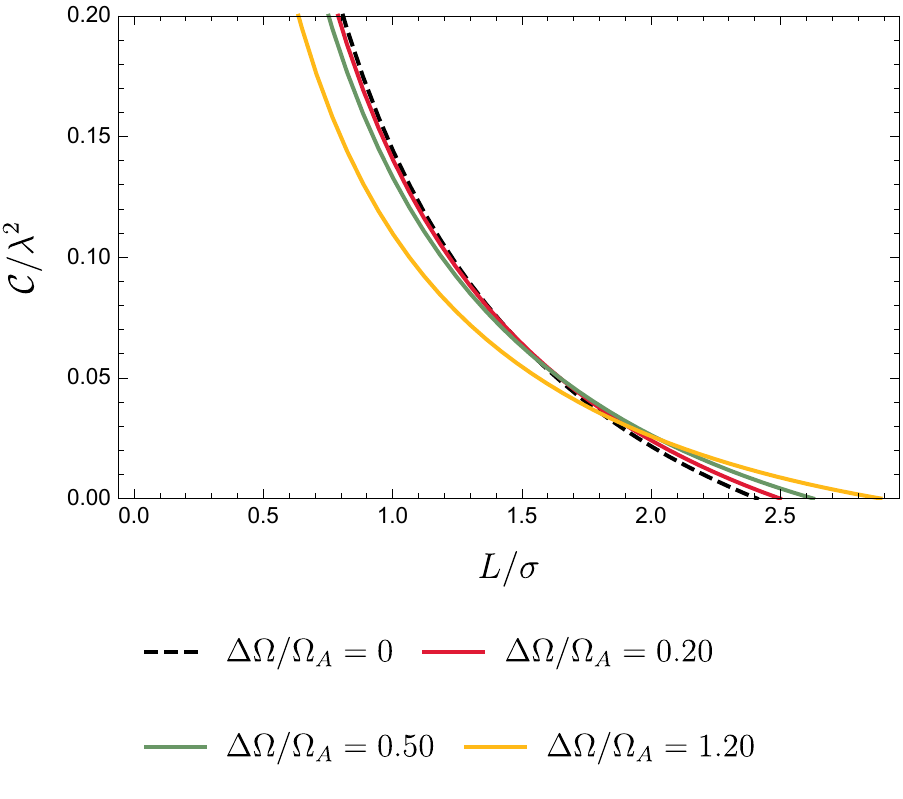}}\qquad
 \subfloat[$\Omega_A\sigma=1.20$]{\label{con-L22}\includegraphics[width=0.47\linewidth]{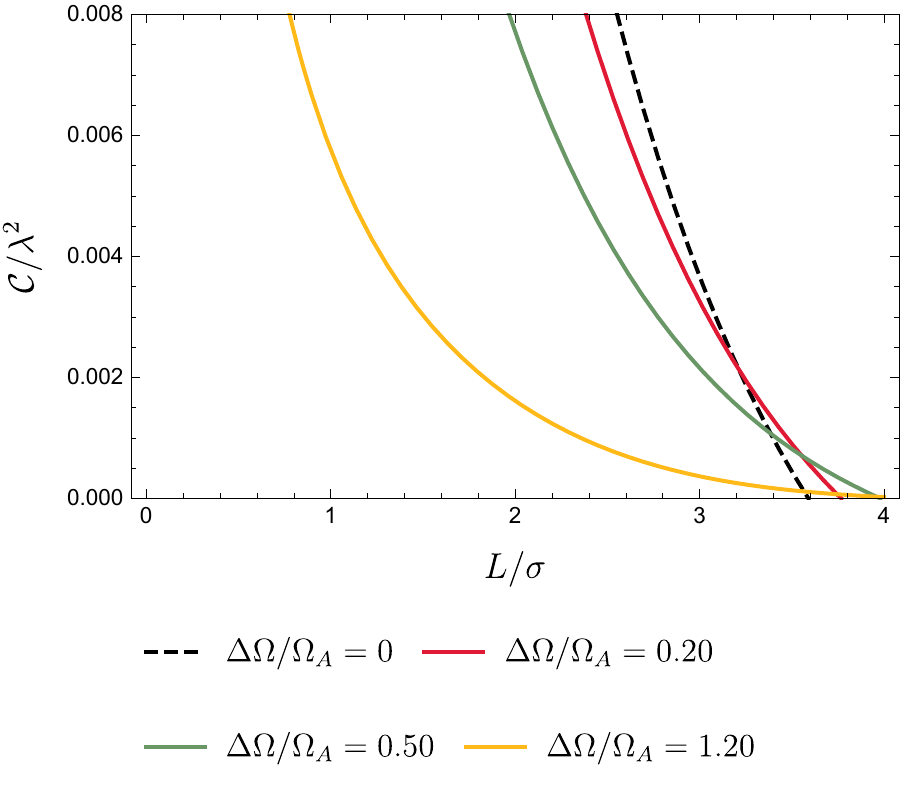}}
\caption{The concurrence is plotted as a function of the separation between the detectors, $L/\sigma$, for $\Delta\Omega/\Omega_A=\{0,0.20,0.50,1.00,1.20\}$ with fixed $\Omega_{A}\sigma=0.50$ in (a) and  $\Omega_{A}\sigma=1.20$ in (b).  Here, all relevant physical quantities are rescaled with the parameter $\sigma$ to be unitless for convenience. }\label{con-L}
 \end{figure}

In Fig.~(\ref{con-L}), the concurrence is plotted as  a function of $L/\sigma$ for various $\Delta\Omega/\Omega_A$.  Remarkably, regardless of the value of $\Delta\Omega/\Omega_A$, the  harvested entanglement decreases as the detector separation increases, which is in accordance with the existing results of entanglement harvesting by identical detectors ($\Delta\Omega=0$) in the literature~\cite{MST:2016,Zhjl:2018,Zhjl:2019,Zhjl:2021,Zhjl:2021-2}. So, the existence of an energy gap difference  does not change the property of concurrence as a monotonically decreasing function of  the detector separation. However, it is noteworthy that the gap difference does impact the amount of the entanglement harvested. % by two identical detectors is not always the most.
As shown in Fig.~(\ref{con-L}),  only for small detector separations ($L/\sigma<1$) could the identical detectors with an equal energy gap  harvest more entanglement,  and an evident crossover when non-identical detectors harvest more entanglement would emerge as the inter-detector separation increases to sufficiently large as compared to the interaction duration parameter $\sigma$. The smaller the gap difference, the sooner the crossover occurs as the separation increases.  This implies that non-identical detectors with an energy gap difference at a sufficiently large separation are bound to harvest more entanglement  via locally interacting with vacuum fields than  identical detectors.  In this sense, the presence of an energy gap difference is conducive to entanglement harvesting. It should be emphasized that the  appearance of the crossover phenomenon is  irrelevant to the value of $\Omega_A$ in the sense that  a  larger  $\Omega_A\sigma$ just pushes the crossover point to a larger inter-detector separation $L/\sigma$,  making the quantitative details of  the crossover point a little different for different values of $\Omega_A\sigma$ (compare Figs.~(\ref{con-L11}) and~(\ref{con-L22})).

We could qualitatively understand the  appearance of the crossover phenomenon  as follows. As can be  seen from Eq.~(\ref{app-c1}),  for a very small inter-detector separation,  the two  detectors with an equal energy gap ($\Delta\Omega=0$)  harvest  more entanglement as factor $e^{-\sigma^{2}(2\Omega_A+\Delta\Omega)^{2}/{4}}$ attains the maximum value  for  $\Delta\Omega=0$.  However,  for not too small $L/\sigma$, the transition probabilities, $P_A$ and $P_B$, cannot be neglected in the concurrence~(\ref{condef}) any more.
We could roughly estimate the derivative of the concurrence with respect to $\Delta\Omega$  as  follows
\begin{align}\label{de-C-1}
\frac{\partial{\cal{C}}{(\rho_{AB})}}{\partial(\Delta\Omega)}&\approx\frac{\lambda^2\sigma}{4\sqrt{\pi}}\sqrt{\frac{P_A}{P_B}}{\rm{Erfc}}[(\Omega_A+\Delta\Omega)\sigma]
-\frac{(\Omega_A+\Delta\Omega)\sigma^4\lambda^2}{L^2\pi}
{e}^{-\frac{\sigma^{2}[\Omega_A^2+(\Omega_A+\Delta\Omega)^{2}]}{2}}\nonumber\\
&\sim\frac{\lambda^2\sigma}{\pi}{e}^{-\frac{\sigma^{2}[\Omega_A^2+(\Omega_A+\Delta\Omega)^{2}]}{2}}\Big\{\frac{\sqrt{\pi}}{4}e^{(\Omega_A+\Delta\Omega)^{2}\sigma^2}
{\rm{Erfc}}[(\Omega_A+\Delta\Omega)\sigma]
-\frac{(\Omega_A+\Delta\Omega)\sigma^3}{L^2}\Big\}\;.
\end{align}
In the last line of Eq.~(\ref{de-C-1}),  the approximation  $P_D\propto{e}^{-\sigma^{2}\Omega_D^2}$, i.e., Eq.~(\ref{PDExpression}),  has been utilized.
If the inter-detector separation is large enough ($L\gg\sigma$), we have  ${\partial{\cal{C}}{(\rho_{AB})}}/{\partial(\Delta\Omega)}>0$. This means that the  concurrence  is now an increasing function of $\Delta\Omega$, which also explains why the non-identical detectors  could harvest more entanglement than the identical detectors in the regime of large inter-detector separations.  %%%%%%Remarkably,

 Moreover, Fig.~(\ref{con-L}) also demonstrates  that the harvesting-achievable separation range  is singularly affected by  the energy gap difference. In general, the identical detectors would have a  relatively  small  harvesting-achievable separation range in comparison with non-identical detectors. More details of this property will be discussed later.

\begin{figure}[!ht]
\centering
\subfloat[$\Omega_A\sigma=0.50$]{\label{con-w11}\includegraphics[width=0.47\linewidth]{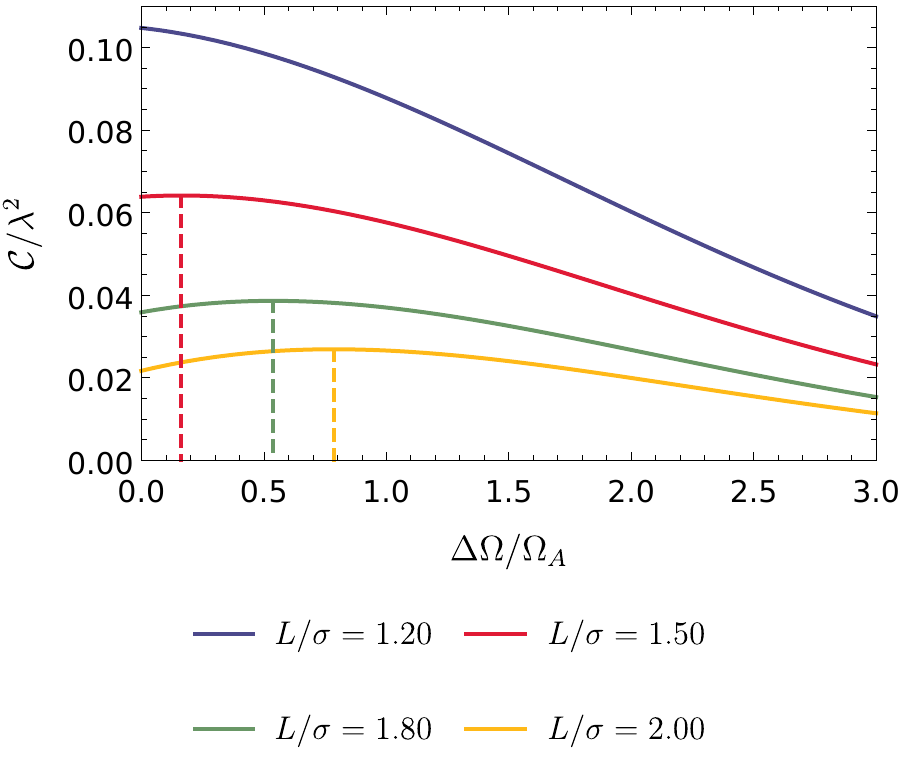}}\quad
 \subfloat[$\Omega_A\sigma=1.20$]{\label{con-w22}\includegraphics[width=0.47\linewidth]{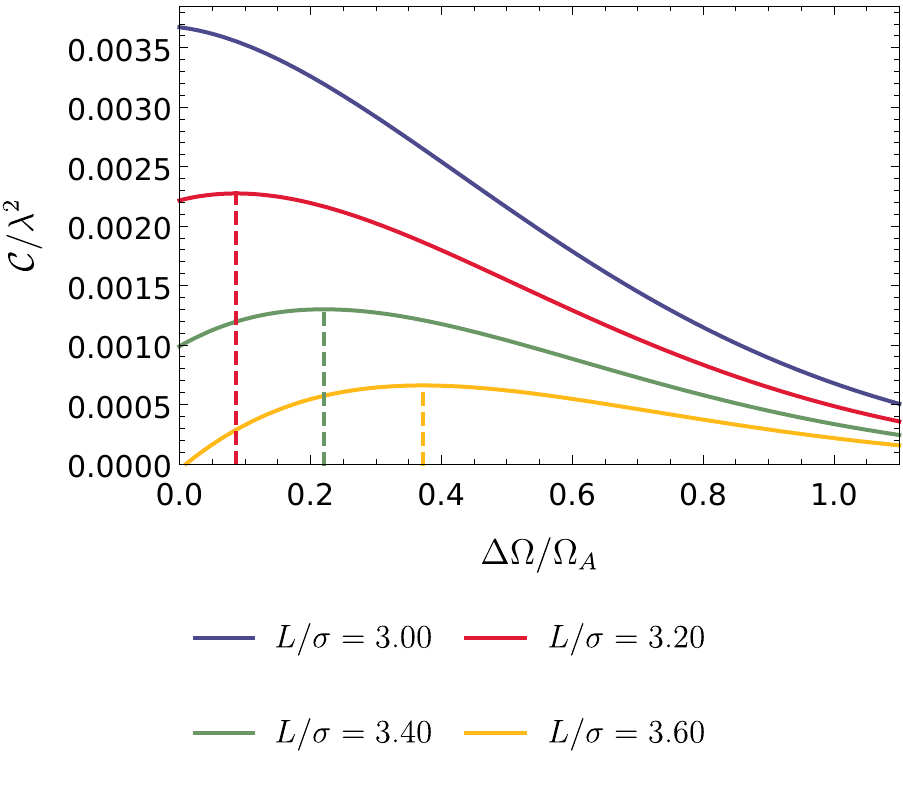}}
\caption{The concurrence versus $\Delta\Omega/\Omega_A$ for various  $L/\sigma$ with $\Omega_{A}\sigma=0.50$ in (a) and $\Omega_{A}\sigma=1.20$ in (b).  The vertical dashed lines in the plots indicate where the peaks are located. }\label{con-w}
 \end{figure}

To gain a better understanding of the influence of an energy gap difference on  entanglement harvesting, we further plot the concurrence versus the energy gap difference in Fig.~(\ref{con-w}). As we can see, the entanglement in general degrades with the increasing energy gap difference $\Delta\Omega$ (in the region of large $\Delta\Omega/\Omega_A$) no matter whether $\Omega_A\sigma$ is small or large, but it remarkably displays a peaking behavior  when the inter-detector separation is sufficiently larger than the duration parameter $\sigma$. This  reveals that there is an optimal value of the energy gap difference (denoted by $\Delta\Omega_p$) that renders the concurrence to peak. % which suggests that the non-identical detectors  are more likely  beneficial to implement entanglement harvesting tasks for not too small detector separation ($L/\sigma>1$).
Moreover, we also find that  the larger the separation,  the larger the optimal value of the energy gap difference, and the larger  $\Omega_A\sigma$,  the larger the inter-detector separation with which the peaking behavior occurs. This property can actually be  deduced from  Eq.~(\ref{de-C-1}). Since $e^{x^2}{\rm{Erfc}}(x)$ is an exponential-like monotonically decreasing function of its argument~\cite{Abramowitz:1964}, a  larger inter-detector separation requires  a larger  $\Omega_A\sigma$ or $\Delta\Omega\sigma$  to make ${\partial{\cal{C}}{(\rho_{AB})}}/{\partial(\Delta\Omega)}=0$ which  governs  where the peak locates.

\begin{figure}[!ht]
  \centering
 \includegraphics[width=0.65\linewidth]{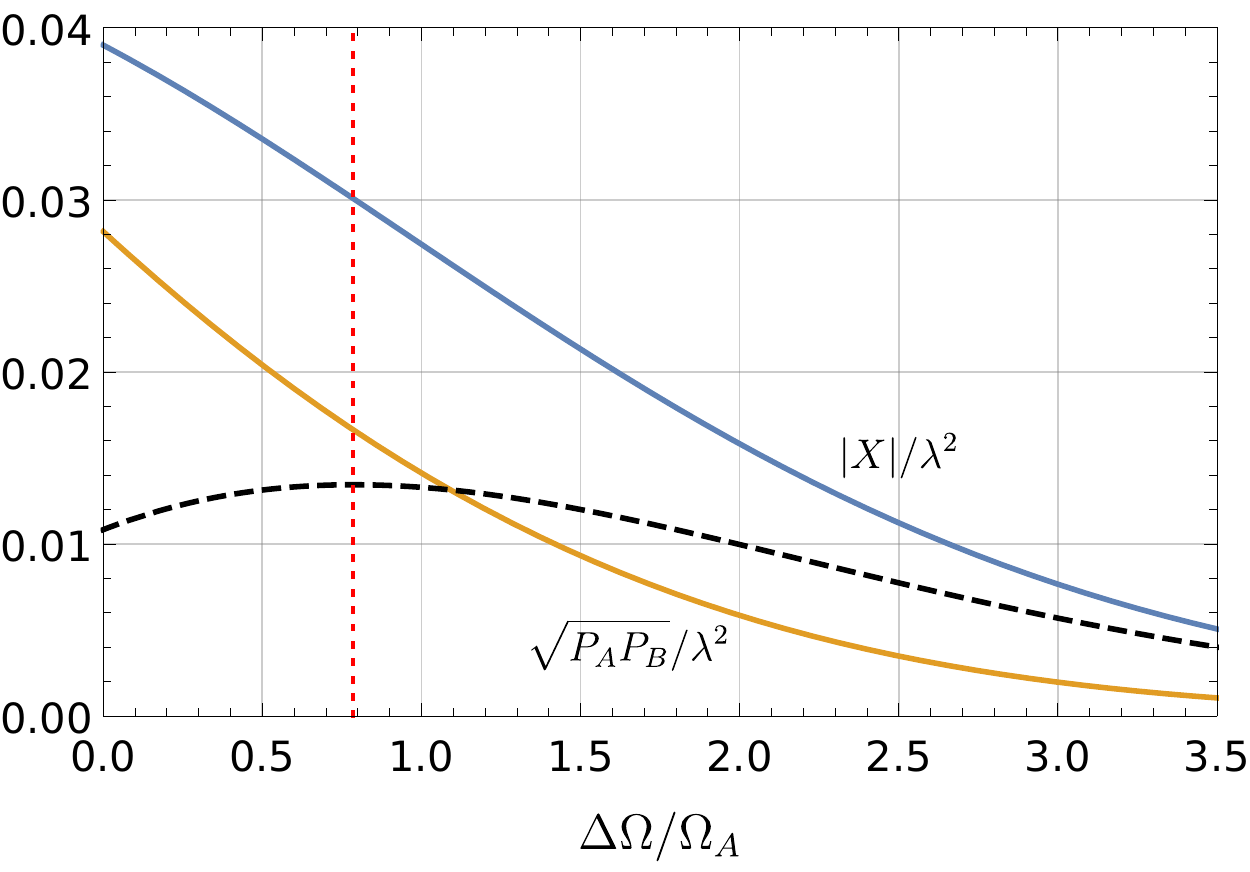}
  \caption{The correlation term $|X|$ and the geometric mean of the detectors' transition probabilities $\sqrt{P_AP_B}$ are plotted as a function of  $\Delta\Omega/\Omega_A$ with $\Omega_A\sigma=0.50$ and $L/\sigma=2.00$. Here, the dashed black line represents the difference between $|X|$ and $\sqrt{P_AP_B}$ , and the vertical dashed line indicates where the peak concurrence occurs, i.e., $|X|-\sqrt{P_AP_B}$ is maximum.}\label{pxvspmega} \end{figure}

\begin{figure}[!ht]
  \centering
 \includegraphics[width=0.7\linewidth]{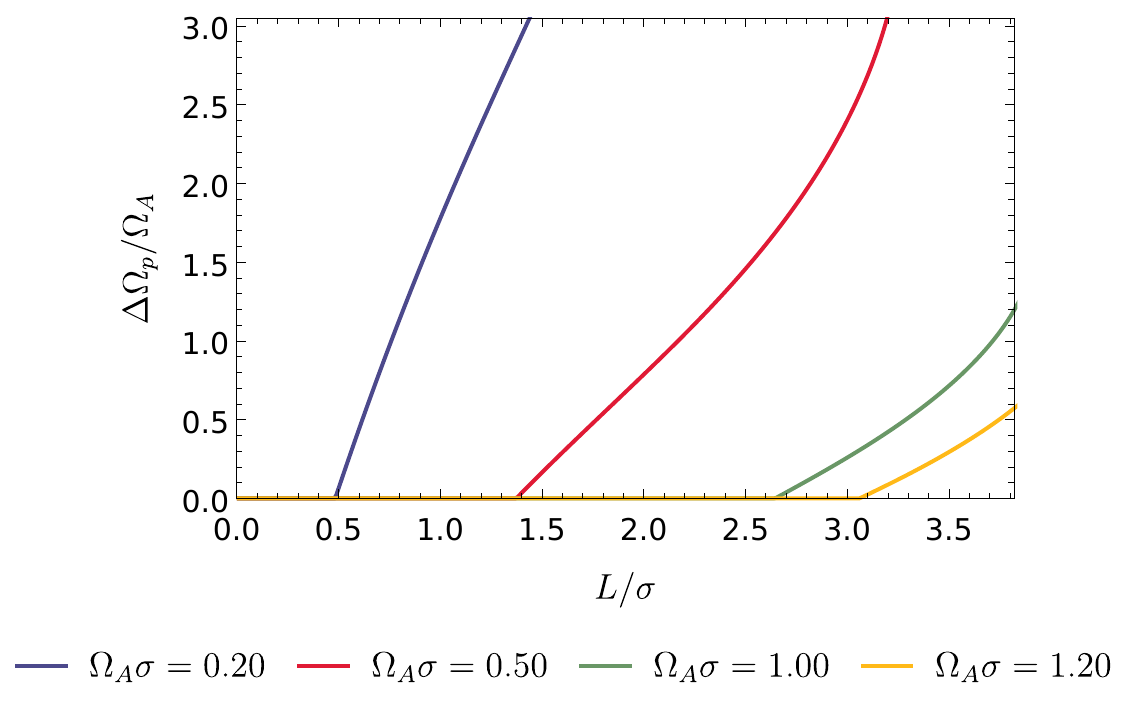}
  \caption{The plot of $\Delta\Omega_{p}/\Omega_A$ versus $L/\sigma$ for various $\Omega_A\sigma=\{0.20,0.50,1.00,1.20\}$. Here, $\Delta\Omega_{p}$ denotes the optimal value of the energy gap difference between the two detectors that peaks the concurrence. It is easy to get that $\Delta\Omega_{p}/\Omega_A$ is a general increasing function  in the region of large enough $L/\sigma$.}\label{wbrwa-L}
 \end{figure}

%%%%%%%%%%%%%%%%%%
In order to analyze the concurrence peaking phenomenon more clearly,  we plot the behaviors of both $|X|$ and $\sqrt{P_AP_B}$
versus $\Delta\Omega/\Omega_A$  in Fig.~(\ref{pxvspmega}). In fact, according to Eq.~(\ref{PDExpression}) and Eq.~(\ref{Expression-X}),  it is easy to see that both the geometric mean of  the transition probabilities $\sqrt{P_AP_B}$ and the  correlation term  $|X|$ decrease as the energy gap difference $\Delta\Omega$ increases. But the rate of decrease  is different. As shown in Fig.~(\ref{pxvspmega}), the difference between $|X|$ and $\sqrt{P_AP_B}$ at the beginning grows with increasing $\Delta\Omega$. However, when the energy gap difference grows across  $\Delta\Omega_p$(indicated by the vertical dashed line in Fig.~(\ref{pxvspmega})), the  correlation term $|X|$ would decrease more rapidly than $\sqrt{P_AP_B}$, resulting in a peak of $|X|-\sqrt{P_AP_B}$ therein. As a consequence, the concurrence would remarkably peak at the optimal value of the energy gap difference $\Delta\Omega_p$. Here, we have plotted $\Delta\Omega_{p}/\Omega_A$ versus $L/\sigma$ in Fig.~(\ref{wbrwa-L}) as a supplement to analyze how the inter-detector separation influences the optimal value of the energy gap difference. %that corresponds to the peak concurrence.
It is easy to discern that the optimal energy gap difference is in general an increasing function of $L/\sigma$ over the region of not too small $L/\sigma$, and a large $\Omega_A\sigma$ renders the  curve of function $\Delta\Omega_p/\Omega_A$ to  move rightward along the axis of $L/\sigma$.

%large $\Omega_A\sigma$ should call for large $L/\sigma$ to make the non-identical detectors harvest the most entanglement.
%%%%%%%%%%%%%%%%%

\begin{figure}[!ht]
  \centering
 \includegraphics[width=0.7\linewidth]{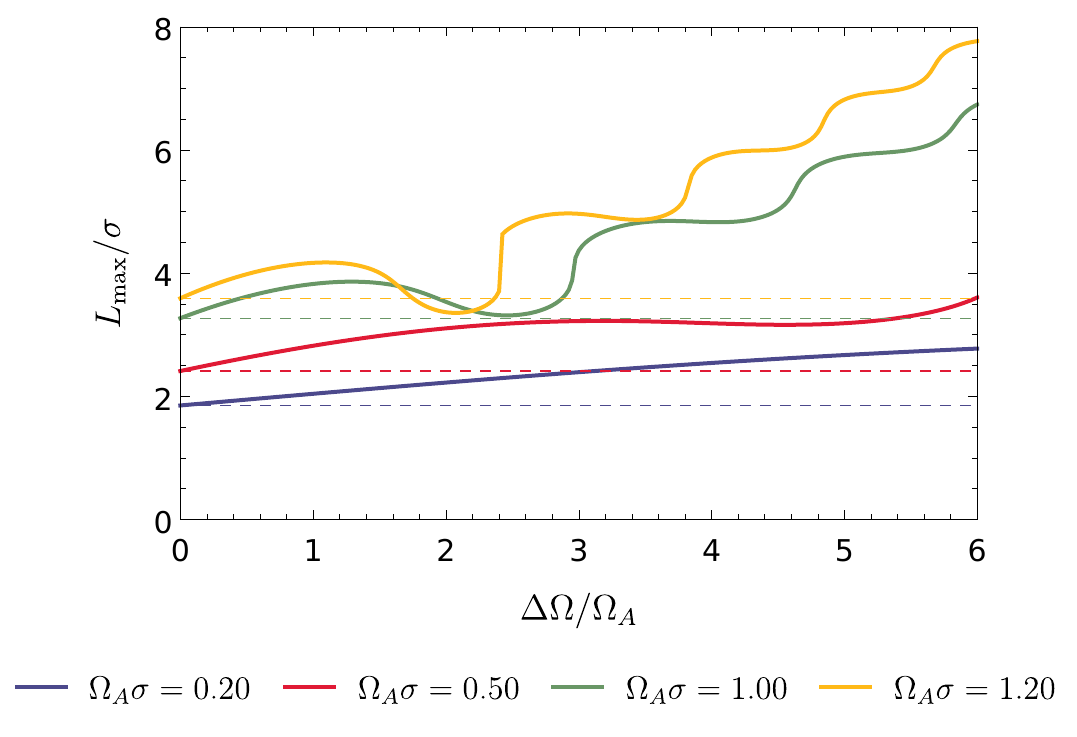}
  \caption{The maximum harvesting-achievable separation, $L_{\rm{max}}/\sigma$, is plotted as a function of $\Delta\Omega/\Omega_A$ for $\Omega_{A}\sigma=\{0.20,0.50,1.00,1.20\}$. The dashed horizon lines indicate the corresponding values of $L_{\rm{max}}/\sigma$ for identical detectors (i.e., $\Delta\Omega=0$). It is easy to see  the oscillatory behavior of $L_{\rm{max}}/\sigma$ would emerge for large $\Omega_A\sigma$.}\label{Lmax-wbrwa}
 \end{figure}

Now, we turn to investigate the influence of an energy gap difference on the harvesting-achievable separation range. As mentioned before, for extremely large energy gaps ($\Omega_A\sigma\gg1$), the energy gap difference does  enlarge the harvesting-achievable range. For not too large energy gaps, the detailed behavior of the harvesting-achievable separation range is further depicted in  Fig.~(\ref{Lmax-wbrwa}). Obviously, the parameter $L_{\rm{max}}$ generally  is an increasing function of $\Delta\Omega$ for small $\Omega_A\sigma$. This means that the non-identical detectors   possess more spacious ``room"  for entanglement extraction, and an energy gap difference between two detectors assists entanglement harvesting  in the sense of harvesting-achievable range.  As is shown in Fig.~(\ref{Lmax-wbrwa}), as the energy gap $\Omega_A$ grows close to and a little beyond $1/\sigma$, the  harvesting-achievable range ($L_{\rm{max}}/\sigma$) overall is  still an  increasing function, but with obvious oscillations  as a result of the oscillatory  cosine  and exponential functions of $\Delta\Omega$ in  Eq.~(\ref{Expression-X}). Such obvious oscillations can even make  the value of $L_{\rm{max}}/\sigma$ have one opportunity to be smaller than that for identical detectors (see $\Omega_A\sigma=1.20$ in Fig.~(\ref{Lmax-wbrwa})).
Hence, the energy gap difference between the two detectors does have significant influence on entanglement harvesting phenomenon, i.e., overall it enlarges the harvesting-achievable separation range and results in a peak of  the  entanglement harvested for a sufficiently large inter-detector separation.

%%%%%%%%%%%

\section{conclusion}
We have performed  a detailed investigation on the entanglement harvesting phenomenon of two non-identical detectors with different energy gaps  locally interacting with massless scalar fields in (3+1)-dimensional Minkowski spacetime.  We analyzed the influence of an energy gap difference between the two detectors on entanglement harvesting from two perspectives: the amount of  entanglement harvested   and the harvesting-achievable separation, mainly focusing on the question whether the non-identical detectors are more easily to get  entangled  than the identical detectors and when the non-identical detectors could harvest more entanglement from the fields in vacuum  during the entanglement harvesting process.

For the perspective of the amount, the harvested entanglement in general degrades with the increasing  separation between the two detectors regardless of the energy gap. However,   the non-identical detectors could extract more entanglement from the vacuum state than the identical detectors if  the inter-detector separation is  not too small with respect to the interaction duration parameter. There seems to be an optimal value of the energy gap difference between the non-identical detectors that renders the  harvested entanglement to  peak if the inter-detector separation is sufficiently large.
In addition, we also demonstrate how the optimal value of energy gap difference depends on the inter-detector separation with the energy gap of one detector ($\Omega_A$) fixed. We find that  the larger the inter-detector separation, the greater the  optimal value of the energy gap difference.

 Regarding the harvesting-achievable range of the inter-detector separation,  we find that the harvesting-achievable range is an increasing function of the energy gap difference when the energy gap of the detector with a smaller gap, $\Omega_A$, is  very large or  small as compared to the Heisenberg  energy $1/\sigma$. However,  when $\Omega_A$ becomes comparable to the Heisenberg energy $1/\sigma$ %(i.e., the energy gap is close to and beyond the detector's Heisenberg energy)
 ,
 the harvesting-achievable range overall increases as the gap difference increases, although with obvious oscillations in the regime of large $\Delta\Omega/\Omega_A$. Therefore we conclude  that the presence of  an energy gap difference generally has a positive influence on the harvesting-achievable range. In other words, the presence of an energy gap difference  is conducive to entanglement harvesting of two detectors in the sense of  harvesting-achievable separation range.

 In summary,  non-identical detectors with different energy gaps could harvest more entanglement from the vacuum state of quantum fields than identical detectors  as long as the inter-detector separation is  not too small. Furthermore, the presence of an energy gap difference can  in general enlarge the harvesting-achievable range of the separation between detectors. Our results illustrate that the energy gap difference is an important physical quantity like those characterizing  spacetime topology, curvature and  detector's motion  to manipulate  entanglement harvesting.

\begin{acknowledgments}
  This work was supported in part by the NSFC under Grants No. 11690034, No.12075084 and No.12175062, and the Research Foundation of Education Bureau of Hunan Province, China under Grant No.20B371.
\end{acknowledgments}

\appendix
\section{The derivation of X}\label{appd1}
To verify Eq. (\ref{Expression-X}), let us begin from Eq. (\ref{xxdef}), which satisfies
 \begin{align}\label{xxdef-2}
 X=&-\lambda^{2} \int_{-\infty}^{\infty} d\tau \int_{-\infty}^{\tau}d \tau' \chi(\tau)\chi(\tau')
\Big[ e^{-i(\Omega_{A}\tau'+\Omega_{B}\tau)}W(x_{A}(\tau'),x_{B}(\tau))+e^{-i(\Omega_{A}\tau+\Omega_{B}\tau')}W(x_{B}(\tau'),x_{A}(\tau))\Big]\nonumber\\
 =&-\lambda^{2}\int_{-\infty}^{\infty}du\chi(u)\chi(u-s)e^{-i(\Omega_{A}+\Omega_{B})u}
 \int_{0}^{\infty}ds\Big[e^{i\Omega_{A}s}W(s)+e^{i\Omega_{B}s}W(s)\Big]\;,
 \end{align}
 where we have introduced $u := \tau$ , $s:=\tau-\tau'$ and considered $W(s)=W(x_{A}(\tau'),x_{B}(\tau))=W(x_{B}(\tau'),x_{A}(\tau))$  due to the fact that two detectors are
at rest with respect to one another.
Carrying out the integration with respect to $u$, we have
  \begin{align}\label{xxdef-3}
 X=&-\lambda^{2}\sqrt{\pi}\sigma e^{-\sigma^{2}(\Omega_{A}+\Omega_{B})^{2}/4}
 \int_{0}^{\infty}ds e^{-\frac{s^2}{4\sigma^2}}\Big[
 e^{i(\Omega_{A}-\Omega_{B})s/2}W(s)+ e^{-i(\Omega_{A}-\Omega_{B})s/2}W(s)\Big]\;.
 \end{align}
 The Wightman function Eq.~(\ref{wightdef}) for two static detectors then can be written as
 \begin{align}\label{Wightf-2}
  W(s)=-\frac{1}{4\pi^{2}}\frac{1}{(-s-i\epsilon)^{2}-L^{2}}\;,
\end{align}
Substituting Eq.~(\ref{Wightf-2}) into Eq.~(\ref{xxdef-3}) yields
\begin{align}\label{xxdef-4}
 X&=\frac{\lambda^{2}\sigma}{4\pi^{3/2}} e^{-\sigma^{2}(2\Omega_A+\Delta\Omega)^{2}/4}
 P\int_{-\infty}^{\infty} \frac{e^{-s^{2}/(4\sigma^{2})}}{s^{2}-L^{2}}e^{-i\Delta\Omega\cdot{s}/2}ds \nonumber \\
 &\;\;\;\;-i\frac{\lambda^{2}\sigma}{4\sqrt{\pi}L}e^{-\frac{\sigma^{4}(2\Omega_A+\Delta\Omega)^{2}+L^{2}}{4\sigma^{2}}}\cos\big(\frac{\Delta\Omega{L}}{2}\big)\;,
\end{align}
where we have defined  $\Delta\Omega:=\Omega_B-\Omega_A$ and utilized the Sokhotski formula
\begin{equation}\label{id0}
\frac{1}{x\pm{i}\epsilon}=P\frac{1}{x}\mp{i}\pi\delta(x)\;.
\end{equation}
Here, the corresponding integration in Eq.~(\ref{xxdef-4}) can be carried out  by using the Fourier transforms, that is
\begin{align}\label{integral-2}
 &P\int_{-\infty}^{\infty} \frac{e^{-s^{2}/(4\sigma^{2})}}{s^{2}-L^{2}}e^{-i\Delta\Omega\cdot{s}/2}ds=P\int_{-\infty}^{\infty} \frac{2e^{-t^{2}/\sigma^2}}{4t^{2}-L^{2}}e^{-i\Delta\Omega\cdot{t}}dt
 \nonumber\\&=\mathcal{F}\Big[2e^{-t^{2}/\sigma^2}\cdot\frac{1}{4t^{2}-L^{2}}\Big]
 =\frac{1}{2\pi} \mathcal{F}\big[2e^{-t^{2}/\sigma^2}\big] \ast \mathcal{F}\big[\frac{1}{4t^{2}-L^{2}}\big] \nonumber \\
 &=-\frac{\pi}{2L}e^{-L^{2}/(4\sigma^{2})}e^{-i\Delta\Omega{L}/2}
 \Big[e^{i\Delta\Omega{L}}{\rm{Erfi}}\Big(\frac{L}{2\sigma}-i\frac{\sigma\Delta\Omega}{2}\Big)+{\rm{Erfi}}\Big(\frac{L}{2\sigma}+i\frac{\sigma\Delta\Omega}{2}\Big)\Big]\;,
\end{align}
where the symbols $``\mathcal{F}"$ and $``\ast"$ stand for the Fourier transform and the convolution of two functions,  respectively.

Substituting Eq. (\ref{integral-2}) into Eq. (\ref{xxdef-4}),  one can straightforwardly obtain $X$ in the terms of error functions
\begin{align}
    X&=-\frac{\lambda^{2}\sigma}{8\sqrt{\pi}L}
    e^{-\frac{\sigma^{4}(2\Omega_A+\Delta\Omega)^{2}+L^{2}}{4\sigma^{2}}}
    \;\Big[e^{i\Delta\Omega{L}/2}
    {\rm{Erfi}}\Big(\frac{L-i\sigma^2\Delta\Omega}{2\sigma}\Big)\nonumber\\
    &+e^{-i\Delta\Omega{L}/2}{\rm{Erfi}}\Big(\frac{L+i\sigma^2\Delta\Omega}{2\sigma}\Big)
    +2i\cos\Big(\frac{\Delta\Omega{L}}{2}\Big)\Big]\;.
\end{align}
%%%%%%%%%%%%%%%%%

\def\ACP{AIP Conf. Proc.}
\def\AIHP{Ann. Inst. Henri. Poincar\'e}
\def\AJP{Amer. J. Phys.}
\def\AM{Ann. Math.}
\def\AP{Ann. Phys. (N.Y.)}
\def\APJ{Astrophys. J.}
\def\ASS{Astrophys. Space Sci.}
\def\ATMP{Adv. Theor. Math, Phys.}
\def\CJP{Can. J. Phys.}
\def\CMP{Commun. Math. Phys.}
\def\CPB{Chin. Phys. B}
\def\CPC{Chin. Phys. C}
\def\CPL{Chin. Phys. Lett.}
\def\CQG{Classcal Quantum Gravity}
\def\CTP{Commun. Theor. Phys.}
\def\EASPS{EAS Publ. Ser.}
\def\EPJC{Eur. Phys.  J. C.}
\def\EPL{Europhys. Lett.}
\def\GRG{Gen. Relativ. Gravit.}
\def\IJGMMP{Int. J. Geom. Methods Mod. Phys.}
\def\IJMPA{Int. J. Mod. Phys. A}
\def\IJMPD{Int. J. Mod. Phys. D}
\def\IJTP{Int. J. Theor. Phys.}
\def\JCAP{J. Cosmol. Astropart. Phys.}
\def\JGP{J. Geom. Phys.}
\def\JETP{J. Exp. Theor. Phys.}
\def\JHEP{J. High Energy Phys.}
\def\JMP{J. Math. Phys. (N.Y.)}
\def\JPA{J. Phys. A}
\def\JPCS{J. Phys. Conf. Ser.}
\def\JPSJ{J. Phys. Soc. Jap.}
\def\LMP{Lett. Math. Phys.}
\def\LNC{Lett. Nuovo Cim.}
\def\MPLA{Mod. Phys. Lett. A}
\def\NPB{Nucl. Phys. B}
\def\PCAM{Proc. Symp. Appl. Math.}
\def\PCPS{Proc. Cambridge Philos. Soc.}
\def\PDU{Phys. Dark Univ.}
\def\PLA{Phys. Lett. A}
\def\PLB{Phys. Lett. B}
\def\PR{Phys. Rev.}
\def\PRA{Phys. Rev. A}
\def\PRD{Phys. Rev. D}
\def\PRE{Phys. Rev. E}
\def\PRL{Phys. Rev. Lett.}
\def\PRX{Phys. Rev. X}
\def\PRSLA{Proc. Roy. Soc. Lond. A}
\def\PTP{Prog. Theor. Phys.}
\def\PRp{Phys. Rept.}
\def\RMP{Rev. Mod. Phys.}
\def\SB{Sci. Bull.}
\def\SPP{Springer Proc. Phys.}
\def\SRTU{Sci. Rep. Tohoku Univ.}
\def\ZPC{Zeit. Phys. Chem.}

%%%%%%%%%%%%%%%%%%%
\end{document}